\documentclass[twocolumn,times,tighten]{aastex631}
\usepackage{amstext}
\usepackage[english]{babel}
\usepackage[utf8x]{inputenc}
\usepackage{textgreek}

\newcommand{\teff}{$T_\mathrm{eff}$}

\newcommand{\logg}{$\log g$}
\newcommand{\feh}{[Fe/H]}

\newcommand{\kms}{km\,s$^{-1}$}

\begin{document}

\title{The First Chemical Census the Milky Way's Nuclear Star Cluster}


\correspondingauthor{Govind Nandakumar}
\email{govind.nandakumar@fysik.lu.se}

\author[0000-0002-6077-2059]{Govind Nandakumar}
\affil{Division of Astrophysics, Department of Physics, Lund University, Box 118, SE-22100 Lund, Sweden}
\affil{Aryabhatta Research Institute of Observational Sciences, Manora Peak, Nainital 263002, India}

\author[0000-0001-6294-3790]{Nils Ryde}
\affil{Division of Astrophysics, Department of Physics, Lund University, Box 118, SE-22100 Lund, Sweden}


\author[0000-0002-6590-1657]{Mathias Schultheis}
\affil{Université Côte d’Azur, Observatoire de la Côte d’Azur, Laboratoire Lagrange, CNRS, Blvd de l’Observatoire, 06304 Nice, France}

\author[0000-0003-0427-8387]{R. Michael Rich}
\affil{Department of Physics and Astronomy, UCLA, 430 Portola Plaza, Box 951547, Los Angeles, CA 90095-1547, USA}

\author{Paola di Matteo}
\affil{GEPI, Observatoire de Paris, PSL Research University, CNRS, Place Jules Janssen, 92195 Meudon, France}

\author[0000-0002-5633-4400]{Brian Thorsbro} 
 \affil{Université Côte d’Azur, Observatoire de la Côte d’Azur, Laboratoire Lagrange, CNRS, Blvd de l’Observatoire, 06304 Nice, France}

 \author[0000-0001-7875-6391]{Gregory Mace}  
 \affil{Department of Astronomy and McDonald Observatory, The University of Texas, Austin, TX 78712, USA}





\begin{abstract}

An important step in understanding the formation and evolution of the Nuclear Star Cluster (NSC) is to investigate its chemistry and chemical evolution. Additionally, exploring the NSC's relationship to the other structures in the Galactic Center and the Milky Way disks is of great interest. Extreme optical extinction has previously prevented optical studies, but near-IR high-resolution spectroscopy is now possible.
Here, we present a detailed chemical abundance analysis of 19 elements - more than four times as many as previously published - for 9 stars in the NSC of the Milky Way, observed with the IGRINS spectrometer on the Gemini South telescope. This study provides new, crucial observational evidence to shed light on the origin of the NSC. We demonstrate that it is possible to probe a variety of nucleosynthetic channels, reflecting different chemical evolution timescales.
Our findings reveal that the NSC trends for the elements F, Mg, Al, Si, S, K, Ca, Ti, Cr, Mn, Co, Ni, Cu, and Zn, as well as the s-process elements Ba, Ce, Nd, and Yb, generally follow the inner-bulge trends within uncertainties. This suggests a likely shared evolutionary history and our results indicate that the NSC population is consistent with the chemical sequence observed in the inner Galaxy (the inner-disk sequence). However, we identify a significant and unexplained difference in the form of higher Na abundances in the NSC compared to the inner-bulge. This is also observed in few Galactic globular clusters, and may suggest a common enrichment process at work in all these systems.

\end{abstract}


\section{Introduction}
\label{sec:intro}

The Nuclear Star Cluster (NSC) of the Milky Way is a unique stellar system  located at the Galactic Center; it is a very compact, massive, and  centrally concentrated  structure with a half-light radius of approximately $4$\,pc
and a stellar mass of $2\times 10^7\,\mathrm M_\odot$ \citep[e.g.][]{schodel:14a,schodel:14b,feld:14,chat:15,fritz:16,feld:17,gallego:20}.
It lies at the center of the much larger Nuclear Stellar Disk (NSD), and Central Molecular Zone (CMZ).

The connection and role of the NSC in the formation and evolution of the Milky Way as a whole remain uncertain. Given that the NSC appears to be a typical nuclear star cluster \citep{neumayer:20}, its connection is likely significant, as suggested by the general scaling relations between the masses, stellar densities, and populations of extragalactic nuclear star clusters and their host galaxies \citep{neumayer:20}. Additionally, scaling relations exist between nuclear star clusters and the supermassive black holes at their centers.

The formation of the NSC is not fully understood, but a few formation scenarios have been proposed \citep[see][for a review]{neumayer:20}. The main scenario is in-situ star formation where, for example, gas is funneled into the central few parsecs, triggering star
formation, driven by several mechanisms, including bar-driven gas infall, dissipative nucleation, tidal compression, or magneto-rotational instabilities. Infall of massive stellar clusters is also an expected process \citep{neumayer:20} and could account for the more metal-poor populations \citep[e.g.,][]{capu:93,mastro:12,Hartmann2011, Arca-Sedda2014}. These scenarios need to be constrained with observations. For example, globular cluster infall cannot explain the presence of young stars \citep[e.g.,][]{Feldmeier-Krause2015} but globular clusters, as they are observed in the Milky Way today, may have contributed a non-negligible fraction \citep[approximately  18\%, according to][]{Dong:2017} of the mass of the NSC. Although unlikely to be the dominant formation mechanism,
this appears significant at low metallicities ($-1.5\lesssim$\feh$ \lesssim -1$), particularly considering that stars with these metallicities do not constitute the majority of NSC population \citep[see, e.g.,][]{Feldmeier-Krause:2020}. 

The general understanding of the stellar populations in the NSC  is that they  appear to be predominately old \citep{matsunaga:11,lara:20,schodel:20,gallego:24}. However, the presence of young stars provides evidence of recent star formation. Especially, within 0.5 pc of the center, a population of  massive young ($<10$ Myrs) stars has formed in situ \citep{neumayer:20}.  According to \citet{schodel:20}, the star-formation history  in the NSC has varied over time, but with the majority of star formation occurring  more than 10 Gyrs ago. An intermediate-age population, approximately 3 Gyrs old,  accounts for less than 15\% of the total star formation \citep{schodel:20,lara:21}. \citet{sanders:24} suggest a close evolutionary link between the NSD and NSC and  also infer old ages, but mainly younger than the main bar/bulge  ($\lesssim8$\,Gyr). On the other hand, \citet{chen:23} identify a dominant, metal-rich component (constituting over 90\% of the stellar mass) in the NSC with an age of around 5 Gyrs. Following a minimum in star formation 1-2 Gyr ago, there may have been an increase in star formation within the past few hundred Myrs \citep{blum:03,pfuhl:11,lara:19,neumayer:20}.

New observational constraints can be derived from the chemical composition of the populations in the NSC. Abundance trends as a function of metallicity for different elements, combined with galactic chemical evolution modeling, can constrain, for instance, the star-formation history, nucleosynthetic processes, and the necessity of gas infall, as well as specific effects due to the unique environment at the Galactic Center \citep[such as strong magnetic fields, dense gas, and high turbulence, see e.g.][]{nsc:turbulence}. Since different groups of elements  trace distinct nucleosynthetic pathways with unique evolutionary timescales \citep[see, e.g.,][]{Matteucci:2021,manea:23}, these trends might vary across stellar populations with differing histories. Consequently, they could illuminate differences and relationships between stellar populations in the NSC and those in the NSD, bulge, thick disk, and thin disk of the Milky Way. The goal is to obtain trends for as many elements as possible, covering a wide range of nucleosynthetic processes.

Very few abundance studies of stars in the NSC have been published to date. 
The first was by \citet{cunha:07}, who focused on the young supergiant population and showed elevated $\alpha$-elements at high metallicities for 10 stars. \citet{Guerco:2022a,guerco:22b} further studied young NSC stars and presented CNO and fluorine abundances.  \citet{do:18} derived abundances of the iron-peak elements V and Sc, and the s-element Y and showed that they were elevated by an order of magnitude, based on very strong spectral lines in two red giants. However, \citet{thorsbro:2018} attributed  these strong lines mostly to line-formation effects in cool giants rather than to extremely high abundances. Later, \citet{thorsbro:2020} determined the first [Si/Fe] for 15 stars and found an enhanced trend at supersolar metallicities. 

This paper is the fourth in a series characterizing the chemistry of the intermediate-age to old red giant populations in the NSC using high-resolution, near-IR spectroscopy. The first paper, \citet{Rich:2017}, identified a broad metallicity distribution ranging from $-0.5<$\feh$<+0.5$\,dex based on 15 M giants.  The second paper, \citet{thorsbro:2018}, examined  the silicon abundance trends versus metallicity for these 15 NSC stars. In the third paper, \citet{ryde:24}, abundance trends for the $\alpha$ abundances of Mg, Si, and Ca were presented for nine NSC giants. In this study, we present abundance measurements for an additional 16 elements in these nine red giants, extending the discussion to elemental abundance trends for groups such as fluorine, odd-Z, iron-peak, and neutron-capture elements. This marks the first time a total of 19 elements - four times more than previously published - have been determined for NSC stars, providing crucial observational evidence to shed new light on the origin of the NSC. Especially, new elements that evolve on different timescale than the commonly used $\alpha$ elements (fast evolution) can give new insights. Examples of such elements with a slower evolution are the slow neutron-capture elements (mainly from AGB stars) and manganese (mainly from Type Ia supernovae, SNe\,Ia).

\section{The Analysis of the Stellar Spectra}
\label{sec:obs}

The observations, data reduction, and properties of the nine  M giant stars in the NSC analyzed here were presented in \citet{ryde:24}. The stars were all shown to be confined to the NSC and have magnitudes of $12<$H$<14$ and $10<$K$<11$. They were observed at $R \sim$ 45,000 spanning the full H and K bands (1.45 - 2.5 $\mu$m) with the Immersion GRating INfrared Spectrograph \citep[IGRINS;][]{Yuk:2010, Wang:2010, Gully:2012, Moon:2012, Park:2014, Jeong:2014} mounted on the Gemini South telescope \citep{Mace:2018} under the programs 
GS-2022A-Q-208, GS-2023A-Q304, and GS-2024A-Q-304. The signal-to-noise ratios (S/N) of the spectra in the K band are larger than 100 whereas they are a factor of 3 lower in the H band. In the present study we will also use a control sample of the same type of stars  consisting of 50 M giants in the solar neighborhood from the study of \citet{Nandakumar:2023}. These were observed as part of the GS-2020B-Q-305 and GS-2021A-Q302 programs. For most of these stars a S/N of well above 100 was achieved both in the H and K bands, and S/N$>50$ for all. 

The stellar parameters are determined based on the method outlined in \citet{Nandakumar:2023}, and are confined to $3350<T_\mathrm{eff}<3850$\,K, $0.3<\log g<1.5$,  $-0.1<\mathrm{[Fe/H]}<0.5$, and $1.8<\xi_\mathrm{micro}<2.7$. Table~2 in \citet{ryde:24} provides the final stellar parameters of the NSC stars.  The control sample of solar neighborhood stars fall within the ranges of $3350<T_\mathrm{eff}<3800$\,K, $0.3<\log g<1.1$,  $-0.9<\mathrm{[Fe/H]}<0.25$, and $1.8<\xi_\mathrm{micro}<2.7$. Table~3 in \citet{Nandakumar:2023}  provides the stellar parameters for these stars. Typical uncertainties are $\pm$100 K in \teff, $\pm$0.2 dex in \logg, $\pm$0.1 dex in \feh, and $\pm$0.1 km s$^{-1}$ in $\xi_\mathrm{micro}$ \citep[for more details see,][]{ryde:24}.  The abundance uncertainties \citep[based on the discussion in][]{Nandakumar:2023} range from 0.05 to 0.15, and are shown as error bars in the Figures below.  In addition, we redetermined the stellar parameters of the seven NSC stars that converged when assuming thin-disk [O/Fe] abundances as an exercise to show the effect of stellar parameters on the derived elemental abundances. This lead to a decrease in the [O/Fe] by 0.1 to 0.15 dex, which in turn results in a decrease in \teff\, by 50 to 100 K, \logg\, by 0.1 to 0.2 dex, \feh\, by 0.0 to 0.05 dex, and $\xi_\mathrm{micro}$ by 0.0 to 0.15 \kms. The resulting differences in stellar parameters are given in Table~\ref{table:parameters_diff}.

We analyse our stellar spectra by synthesizing modelled spectra generated using the Spectroscopy Made Easy (SME) tool \citep[SME;][]{sme,sme_code}, which calculates the spherical radiative transfer through a relevant stellar atmosphere model defined by its fundamental stellar parameters. The model atmosphere is selected by interpolating within a grid of one-dimensional (1D) Model Atmospheres in a Radiative and Convective Scheme (MARCS) stellar atmosphere models \citep{marcs:08}. 

\citet{Nandakumar:2023,Nandakumar:24_21elements} provide a detailed abundance analysis for our comparison sample of the solar neighbourhood M giants. Based on this investigation we scan all available lines for every element in the H and K bands and carefully select unblended spectral lines. Systematic effects, such as unknown blending lines, can impact certain spectral lines in specific regions of the stellar parameter space and should be avoided. 
Compared to the lines used in \citet{Nandakumar:24_21elements,nandakumar:24}, we have added and removed certain lines for specific elements while determining abundances for NSC stars as well as the comparison sample (solar neighborhood and inner-bulge stellar populations). Specifically, one line (21452.00 \AA) for sodium and one line (22701.00 \AA) for aluminum have been included after confirming that the trends from each line are consistent for the solar neighborhood M giants.
We have removed two lines (15860.21 \AA, 17708.73 \AA) for chromium and one line (21945.50 Å) for nickel due to the larger scatter observed in the NSC trends, likely caused by lower S/N. Furthermore, we removed one aluminum line (21208.18 \AA) and two neodymium lines (15368.14 \AA, 16262.04 \AA) due to strong atomic and molecular blends for NSC stars.



When determining the abundances from every one of these chosen lines, we  define specific local continua around the spectral lines to ensure the best abundance determination. 
The atomic line data are taken from the studies of \citet{Nandakumar:2023,Nandakumar:24_21elements} and the molecular line data are adopted from the line lists of \citet{li:2015}, \citet{brooke:2016},  \citet{sneden:2014}, and \citet{exomol_h2o}  for the CO, CN, OH, and H$_2$O, respectively.
In our analysis, we incorporate departure coefficients from non-local thermodynamic equilibrium (non-LTE) grids for the elements C, N, O, Na, Al, Mg, Si, Ca, Ti, Mn, Fe, Cu, and Ba \citep{amarsi:20,amarsi:22}.

\begin{table*}
\caption{Differences in stellar parameters, C, N, and O abundances for the NSC giants between actual values and when determined with the assumption that oxygen follows thin disk abundance trend. }\label{table:parameters_diff}
\begin{tabular}{c r c r c r r r}
\hline
\hline
 Star & T$_\mathrm{eff}$ & $\log g$  & [Fe/H]  &  $\xi_\mathrm{micro}$ & [C/Fe]  & [N/Fe]  & [O/Fe] \\
 \hline
  & K & log(cm/s$^{2}$) & dex & Km/s & dex  & dex  & dex\\
  \hline
FK48  &  81  &  0.14  &  0.02  &  0.05  &  0.12  &  0.04  &  0.13   \\ 
FK5020265  &  75  &   0.12  &   0.0  &   0.1  &   0.12  &   0.04  &   0.13    \\ 
FK87  &  102  &    0.15  &    -0.01  &    0.16  &    0.12  &    0.06  &    0.14   \\ 
Feld31  &  80  &   0.18  &   0.05  &   0.08  &   0.08  &   0.08  &   0.1    \\ 
Feld84  &  90  &  0.18  &  0.03  &  0.0  &  0.08  &  0.05  &  0.09  \\ 
GC15540  &  70  & 0.14  &  0.03  &  0.15  &  0.11  &  0.07  &  0.13   \\ 
GC16890  &  53  &  0.12  &  0.05  &  0.08  &  0.1  &  0.05  &  0.11    \\ 
\hline

\hline
\hline
\end{tabular}
 
\end{table*}


 \begin{figure*}
  \includegraphics[width=\textwidth]{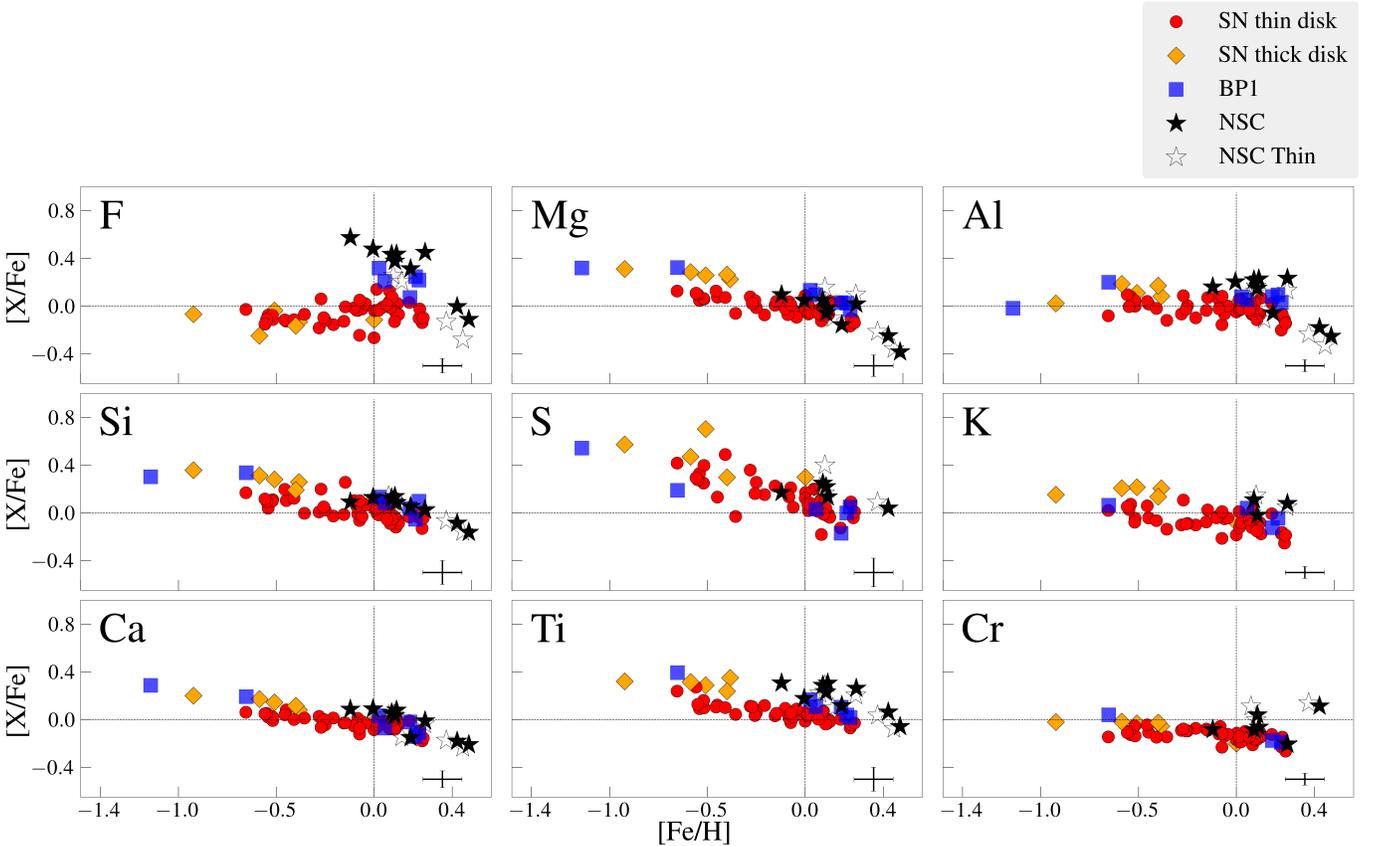}
  \caption{Abundance ratios versus metallicities for F, Mg, Al, Si, S, K, Ca, Ti, and Cr  for the NSC stars (black), inner-bulge stars located 1 degree North of the Galactic Center (blue), thick-disk (orange), and thin-disk stars (red). Black open star symbols represent the abundance ratios versus metallicities for the NSC giants using the stellar parameters determined with the assumption that oxygen abundances follow thin-disk trend (see Table\ref{table:parameters_diff}).} 
  \label{fig:all_trend}%
\end{figure*}

 \begin{figure*}
  \includegraphics[width=\textwidth]{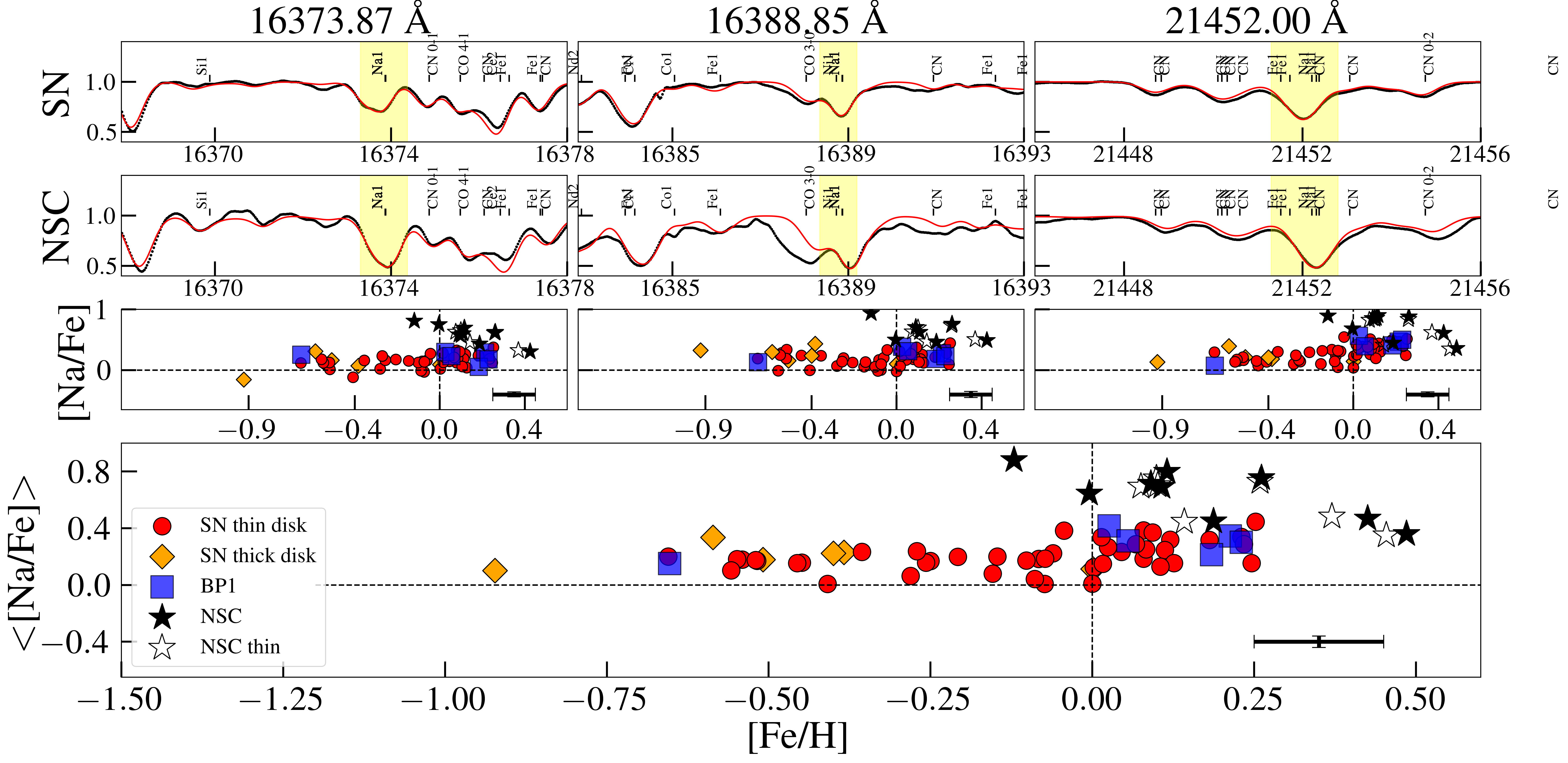}
  \caption{[Na/Fe] versus [Fe/H] for different stellar populations. The Nuclear Star Cluster (NSC) stars are represented by black stars, the inner bulge stars from \citet{nandakumar:24} by blue squares, and the solar neighborhood thin-disk stars are depicted by red filled circles, while the thick-disk stars are shown as orange diamonds. Sodium abundances for NSC stars derived using stellar parameters determined with the thin disk oxygen abundance trend are represented by black open stars. In the upper two panels, the spectral lines used for the analysis are displayed, with a typical solar neighborhood star shown above and a typical NSC star (FK5020265) shown below. The trends derived from each individual spectral line are presented, along with the mean trend, which is displayed in the largest, bottom panel.}
  \label{fig:na_trend}%
\end{figure*}

 \begin{figure*}
  \includegraphics[width=\textwidth]{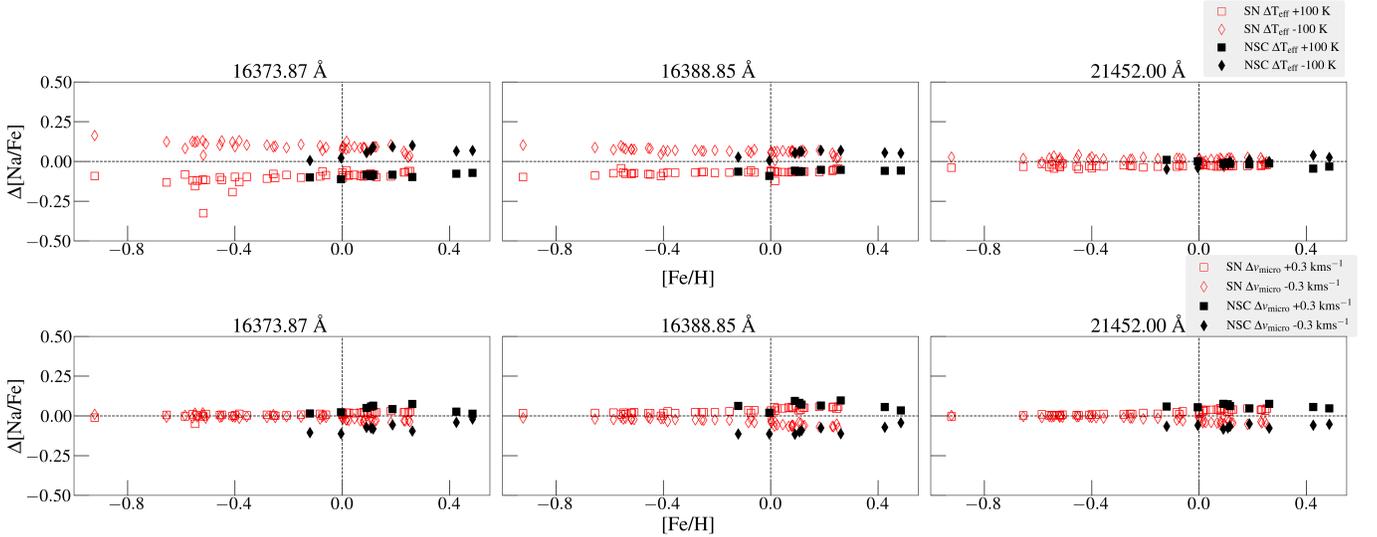}
  \caption{Line-by-line sodium abundance sensitivity to \teff\, (top row) and the microturbulence ($\xi_\mathrm{micro}$; bottom row). Black filled squares and diamonds represent abundance differences corresponding to $\Delta$\teff=+100 K, $\Delta$$\xi_\mathrm{micro}$=+0.3 kms$^{-1}$ and $\Delta$\teff=-100 K, $\Delta$$\xi_\mathrm{micro}$=-0.3 kms$^{-1}$ respectively for the NSC stars. Red open squares and diamonds represent the corresponding abundance differences 
  for the solar neighborhood stars.}
  \label{fig:na_sensitivity_trend}%
\end{figure*}

\section{Results and Discussion}
\label{sec:results}

To identify differences that may reflect distinct chemical evolution in the NSC compared to the solar vicinity, it is essential to minimize systematic uncertainties. Therefore, we will compare the abundance trends of 19 elements, noting that Mg, Si, and Ca were already presented in \citet{ryde:24}, for our  M-giants in the NSC with those derived from the control sample of 50 solar-neighborhood M giants \citep{Nandakumar:2023,Nandakumar:24_21elements}, as well as those from the inner-bulge population investigated by \citet{nandakumar:24}, located $1^\circ$ North of the Galactic Center.
These comparison samples were observed using the same observational setup and analyzed with the same methodology, including an identical stellar parameter scale and, in general, identical spectral lines, as employed in this study. This consistent approach enables direct and differential comparisons, element-by-element and line-by-line. By differentially comparing the detailed abundances of M giants across all stellar populations, we ensure a homogeneous analysis and that any systematic  uncertainties impact all three samples in a similar manner. 

\begin{figure*}
  \includegraphics[width=\textwidth]{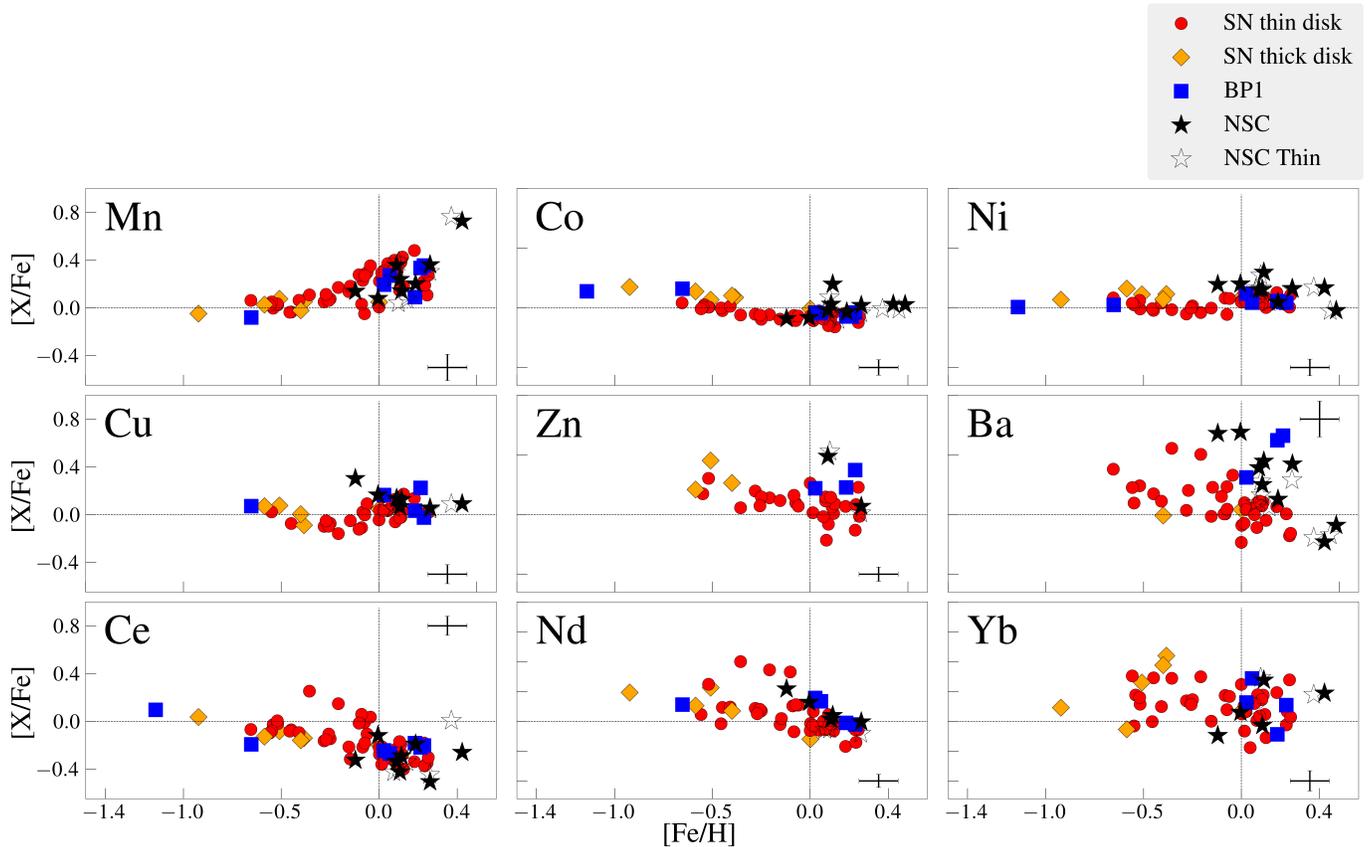}
  \caption{Abundance ratios versus metallicities for Mn, Co, Ni, Cu, Zn, Ba, Ce, Nd, and Yb. Markers are the same as in Figure \ref{fig:all_trend}.}
  \label{fig:heavy_trend}%
\end{figure*}

Our results are shown in Figures \ref{fig:all_trend}, \ref{fig:na_trend}, and \ref{fig:heavy_trend} as abundance ratios plotted against metallicities. The NSC stars are represented by black star symbols, while low-$\alpha$ (thin-disk) stars in the solar neighborhood are shown as red circles, high-$\alpha$ (thick-disk) stars as orange diamonds, and inner-bulge stars as blue squares. Black open stars represent the abundance ratios versus metallicities for the NSC giants derived using the stellar parameters determined with the assumption that oxygen abundances follow thin-disk trend. We note that the variation in these abundances are within the uncertainties marked as error bars in each panel. As discussed in \citet{nandakumar:24}, the inner-bulge stars exhibit a trend that follows {\it the inner-disk sequence} \citep{Haywood:2013,Dimatteo:2016}, following the high-[$\alpha$/Fe] envelope of the metal-rich, thin-disk population in the solar vicinity.

The stars in the NSC exhibit trends that are very similar,  within uncertainties,  to those of the inner-bulge stars for elements following typical $\alpha$-like trends, reflecting rapid chemical evolution driven by massive stars. This is shown by the trend of the mean of Mg, Si, and Ca \citep[Figure 7 in][]{ryde:24}, as well as their individual trends and the sulfur trend shown in Figure \ref{fig:all_trend}. This is further shown in  Figure \ref{fig:all_trend} by the trends of elements like Al \citep[for a nucleosynthetic discussion see, e.g.,][]{matteucci:20}, and  K \citep[e.g.][]{shimansky:03,matteucci:20}. In the case of Ti \citep[e.g.][]{sneden:16} we note a large scatter but within uncertainties, the lines being quite \teff\  sensitive \citep[see][]{Nandakumar:2023}. 


\citet{Nandakumar:2023b} report a very slight fluorine enhancement at super-solar metallicities for stars in the solar neighborhood. Using the HF lines recommended in that study, we find that the F abundances in the NSC stars are only marginally larger than the enhanced values observed in the inner-bulge stars \citep[][see our Figure \ref{fig:all_trend}]{Nandakumar:24_21elements}. \citet{guerco:22b} also find similarly enhanced F abundances in their young NSC supergiants, indicating a comparable enhancement in both the old and young populations. The nucleosynthetic contributions at these metallicities remain under debate \citep[see, e.g.,][]{guerco:22b,Nandakumar:2023b}.  Fluorine determinations from HF lines are highly temperature-sensitive as also indicated by the larger variation in the fluorine abundances derived by varying the stellar parameters (open black stars), and further investigations with larger samples of stars in the NSC is needed to constrain the difference between the populations in the inner Galaxy.

The only element investigated in this study that is significantly different in the NSC compared to the  disk and inner-bulge trends (which are similar) is sodium (see Figure \ref{fig:na_trend}). The ratios are at least a factor of two higher in the NSC stars. The Na lines are well modeled including accurate non-LTE corrections \citep{amarsi:20}, and the lines are neither heavily blended nor saturated, which gives us confidence in the derived abundances. Figure~\ref{fig:na_sensitivity_trend} shows the line-by-line Na abundance sensitivity to \teff\, (top row) and $\xi_\mathrm{micro}$ (bottom row) for both NSC stars (black filled squares and diamonds) and solar neighborhood stars (red open squares and diamonds). The minimal variation ($<$0.1 dex) in [Na/Fe] corresponding to changes of $\pm$100 K in \teff\, and $\pm$ 0.3 kms$^{-1}$ in $\xi_\mathrm{micro}$ demonstrates that these lines are neither significantly \teff-sensitive nor saturated. Similarly, a variation of $\pm$ 0.2 dex in \logg\, resulted in a maximum of 0.1 dex change in [Na/Fe]. Consequently, the derived sodium abundances from these lines can be considered robust and reliable. We also remind that our study is differential among the stars in the different populations, with the only difference being the slightly lower S/N in the NSC giants. Interestingly, \citet{liller1:24} report very similar Na enhancements for the super-solar populations in Liller\,1, which is a  complex, multi-population stellar system in the Galactic bulge. 
It is also interesting to note that \citet{munoz:2018} report a Na spread, but no Na-O anti-correlation, in NGC 6528 -- a metal-rich globular cluster \citep{carretta:01}. Additionally, they do not find a significant Mg or Al spread, which is consistent with the findings of our study. Further theoretical investigations into the different Na sources, such as those in \citet{kobayashi:20}, are needed to explore the cause and implications of these high Na abundances.

The iron-peak elements Cr (Figure \ref{fig:all_trend}), as well as Co and Ni (Figure \ref{fig:heavy_trend}) show flat trends with metallicity that are very similar across all populations. These elements  are formed in comparable amounts by massive stars 
and SNe\,Ia \citep{clayton:2003, kobayashi:11}. The combined contributions from these sources result in a flat trend \citep{kobayashi_nakasato:11,kobayashi:11,lomaeva:19}. 

Mn is overproduced relative to Fe by SNe\,Ia, and is expected to increase with metallicity, reflecting the delayed contribution from SNe Ia, potentially influenced by metallicity-dependent yields \citep{kobayashi_nakasato:11}. Proposed models, such as deflagration or delayed detonation (DDT) in Chandrasekhar-mass SNe Ia \citep{kobayashi:20},  
contribute on a range of timescales, from short to delayed periods ($>1$\,Gyr). 
Consequently,  Mn serves as an effective probe of SNe\,Ia binary systems and delayed nucleosynthesis \citep{kobayashi:20}. The NSC stars, shown in Figure \ref{fig:heavy_trend}, follow the inner-bulge stars well and at the highest metallicities the trend continues to increase. This implies that the binary fraction in the NSC cannot have differed significantly from that in the inner-bulge, despite the much higher stellar densities in the NSC.

Cu seems to be mostly synthesized in the weak-s-process in massive stars as a secondary element (metallicity dependent), but models fail to explain the increase at supersolar metallicities, indicating a later contribution \citep{baratella:21,DelgadoMena:2017}. There might even be some SNe\,Ia and later AGB contributions \citep{kobayashi:11} to its origin. Here too, the NSC stars closely follow the trend of the inner-bulge stars, as seen in Figure \ref{fig:heavy_trend}.

Zn is a transition element between the iron-peak and neutron-capture elements synthesized by the weak s-process, but also through radioactive decay from products synthesized in explosive nucleosynthesis in core-collapse supernovae \citep[with the need of hypernovae;][see Figure \ref{fig:heavy_trend}]{clayton:2003,Kobayashi:2006}. Our zinc abundances are derived from a single weak H-band line, and only two NSC stars have reliable abundance estimates. While these estimates exhibit a large dispersion, they remain consistent with measurements from the inner bulge.

To summarize, the NSC trends for the elements Mg, Al, Si, S, K, Ca, Ti, Cr, Mn, Co, Ni, Cu, and Zn generally follow the inner-bulge trends within uncertainties. The F trend may be slightly higher than the inner-bulge trends, but still consistent within the uncertainties and additional  data is needed to confirm this. 

Of great interest are the main s-process elements 
Ba (s/r=90/10), Ce (s/r=85/15), Nd (s/r=60/40), and Yb (s/r=40/60) \citep[the solar ratio is from][but can vary over time and with metallicity]{bisterzo:14}, as they represent longer timescales due to their synthesis in low- to intermediate-mass AGB stars. These stars release their products with a time delay compared to Type II SNe.  
The NSC trends for Ce and Nd follow  the expected inner-bulge trends, see Figure \ref{fig:heavy_trend}. Unfortunately, Ba \citep{nandakumar:24} and Yb \citep{montelius:22} are challenging to measure due to their positions in the wings of strong CO blends, leading to significant scatter in the data. Nevertheless, they remain consistent with the inner-bulge trends. We note that the slightly higher variation in Ce abundance for the most metal-rich star in the exercise when varying the stellar parameters is possibly due to the use of only one blended line of the four possible Ce lines for this star. 


All these neutron-capture elements have contributions from both the s- and r-processes. Our trends clearly indicate that the s-process contribution is needed, as relying solely on the r-process would fail to explain the observed trends, due to the different time scales for the production of the s- (long) and r-process (short). Yb, being the closest to an r-process element, shows a behavior very similar to that observed in the disk trends, within the scatter.



\section{Conclusions}
\label{sec:conclusion}

We present a detailed chemical abundances analysis of 19 elements for 9 stars in the NSC of the Milky Way. The analysis is based on high-resolution, near-infrared spectra obtained with the IGRINS spectrometer mounted on the Gemini South telescope, covering the full H and K bands. Our aim is to explore the NSC's relation to the inner regions as well as the disks of the Milky Way, which will have implications for the understanding of the formation history and evolution of the Milky Way's central stellar populations. This chemical census of the NSC is the first attempt of its kind and demonstrates that it is possible to probe a number of nucleosynthetic channels, which reflect different chemical evolution timescales, influenced by factors such as star-formation rate and gas infall. Additional data will be required to reduce random uncertainties and further validate our findings.

The s-process elements, along  with those with  contributions from SNe\,Ia, (such as Mn), provide new constraints on the chemical evolution of the NSC, as they differ significantly from the rapidly evolved $\alpha$ elements from SNe Type II.  To summarize our findings, the NSC trends for the elements Mg, Al, Si, S, K, Ca, Ti, Cr, Mn, Co, Ni, Cu, and Zn generally follow the inner-bulge trends within uncertainties. This is also true for the s-process elements. 
Thus, our chemical census of the NSC, here only probing the old- to intermediate-age, metal-rich stars, reveals that the chemical trends versus metallicity  are mostly similar to those of the inner-bulge populations and  likely reflect a shared  evolutionary history. For fluorine a slight enhancement is observed compared to inner bulge trends but similar to that observed for young NSC supergiants.

A clear exception to the similarity is found in the Na abundances, where we observe a significant but unexplained enhancement compared to Na abundances in the inner bulge. These elevated abundances are also detected in the complex, multi-population stellar system in the Galactic bulge, Liller 1, which is chemically similar to Terzan 5 \citep{massari:14}. These systems have been proposed to be fossil fragments from the epoch of Galactic bulge formation \citep{liller1:24}. Our broad metallicity distribution \citep[cf.][]{Rich:2017}, elevated Na abundances, and similar Al and Mg abundances \citep[as in Liller 1;][]{fanelli:24} suggest a common origin or a similar formation history for the NSC and these multi-population systems.


Our findings suggest that the NSC population  is consistent with the chemical sequence observed  in the inner Galaxy (the inner-disk sequence). Given the strong link between the star-formation history  of the thick disk and the bulge in general \citep[see, e.g.,][]{haywood:18}, our results imply that the early star-formation history, that is probed by the abundances of the stars studied here, is similar across scales, ranging from the several-kiloparsec scale of the thick disk to the few-kiloparsec scale of the bulge, and extending to the innermost regions of the Milky Way, in the NSC. This also means that the chemical properties of extragalactic NSCs in galaxies of masses similar to the Milky Way could serve as valuable proxies, at least in part, for understanding the chemical evolution of their host galaxies. This concept was proposed by \citet{pagnini:24} in their discussion of $\omega$  Cen and M54 in the Milky Way, which are thought to be stripped nuclear star clusters of dwarf galaxies \citep{neumayer:20}. Notably, the chemical abundances of the Sagittarius dwarf galaxy and M54, its stripped NSC, show significant similarities \citep{pagnini:24}. 


\begin{acknowledgements}
 G.N.\ acknowledges the support from the Royal Swedish Academy of Sciences (Vetenskapsakademiens stiftelser). N.R.\ acknowledge support from the Swedish Research Council (grant 2023-04744) and the Royal Physiographic Society in Lund through the Stiftelsen Walter Gyllenbergs, Märta och Erik Holmbergs, and Henry och Gerda Dunkers donations. This work used The Immersion Grating Infrared Spectrometer (IGRINS) was developed under a collaboration between the University of Texas at Austin and the Korea Astronomy and Space Science Institute (KASI) with the financial support of the US National Science Foundation under grants AST-1229522, AST-1702267 and AST-1908892, McDonald Observatory of the University of Texas at Austin, the Korean GMT Project of KASI, the Mt. Cuba Astronomical Foundation and Gemini Observatory.
This work is based on observations obtained at the international Gemini Observatory, a program of NSF’s NOIRLab, which is managed by the Association of Universities for Research in Astronomy (AURA) under a cooperative agreement with the National Science Foundation on behalf of the Gemini Observatory partnership: the National Science Foundation (United States), National Research Council (Canada), Agencia Nacional de Investigaci\'{o}n y Desarrollo (Chile), Ministerio de Ciencia, Tecnolog\'{i}a e Innovaci\'{o}n (Argentina), Minist\'{e}rio da Ci\^{e}ncia, Tecnologia, Inova\c{c}\~{o}es e Comunica\c{c}\~{o}es (Brazil), and Korea Astronomy and Space Science Institute (Republic of Korea).
The following software and programming languages made this
research possible: TOPCAT (version 4.6; \citealt{topcat}); Python (version 3.8) and its packages ASTROPY (version 5.0; \citealt{astropy}), SCIPY \citep{scipy}, MATPLOTLIB \citep{matplotlib} and NUMPY \citep{numpy}.
\end{acknowledgements}

%
%


\bibliography{references}{}
\bibliographystyle{aasjournal}






\end{document}